\documentstyle[pre,aps,eqsecnum]{revtex}

\input{epsf.sty}

\newcommand{\figlabel}[1]{\label{fig:#1}} 
\newcommand{\fig}[1]{Fig.~\ref{fig:#1}}
\newcommand{\sectlabel}[1]{\label{sect:#1}}
\newcommand{\sect}[1]{Sec.~\ref{sect:#1}}
\newcommand{\eqnlabel}[1]{\label{eqn:#1}}
\newcommand{\eqn}[1]{Eq.~(\ref{eqn:#1})}
\newcommand{\tablabel}[1]{\label{tab:#1}}
\newcommand{\tab}[1]{Table~\ref{tab:#1}}

\begin{document}

\title{Dynamic and static properties of the invaded cluster algorithm}
\author{K. Moriarty and J. Machta}
\address{Department of Physics and Astronomy,
University of Massachusetts,
Amherst, MA 01003-3720}
\author{L. Y. Chayes}
\address{Department of Mathematics,
University of California,
Los Angeles, CA 90095-1555}

\date{\today}

\maketitle

\begin{abstract}

Simulations of the two-dimensional Ising and 3-state Potts models at 
their critical points are performed using the invaded cluster (IC) 
algorithm.  It is argued that observables measured on a sub-lattice of 
size $l$ should exhibit a crossover to Swendsen-Wang (SW) behavior for 
$l$ sufficiently less than the lattice size $L$, and a scaling form is 
proposed to describe the crossover phenomenon.  It is found that the 
energy autocorrelation time $\tau_{\varepsilon}(l,L)$ for an 
$l\times l$ sub-lattice attains a maximum in the crossover region, and 
a dynamic exponent $z^{{\rm IC}}$ for the IC algorithm is 
defined according to $\tau_{\varepsilon,{\rm max}}\sim 
L^{z^{{\rm IC}}}$.  Simulation results for the 3-state 
model yield $z^{{\rm IC}}=.346\pm .002$ which is smaller 
than values of the dynamic exponent found for the SW and Wolff algorithms and
also less  than the Li-Sokal bound.  The results are less conclusive for the 
Ising model, but it appears that $z^{{\rm IC}}<.21$ and 
possibly that $\tau_{\varepsilon,{\rm max}}\sim \log L$ so that 
$z^{{\rm IC}}=0$---similar to previous results for the SW and Wolff 
algorithms.

\end{abstract} 



\section{Introduction}
Monte Carlo (MC) methods used to simulate classical 
spin systems, such as Potts models, fall primarily into two broad 
classes:  local-update algorithms and cluster algorithms.  Algorithms 
with local update rules, such as the Metropolis algorithm, 
provide an efficient means of simulating these spin systems in 
non-critical regions.  Near a second-order phase transition, however, 
where long-range correlations are present, relaxation times increase 
rapidly with system size.  This phenomenon, known as critical slowing 
down, may be characterized by a dynamic exponent $z$ according to 
$\tau\sim L^{z}$, where $\tau$ is the autocorrelation time measured at 
criticality (roughly, 
the time necessary to generate a statistically independent 
configuration) and $L$ is the system size.  Local-update algorithms 
typically have values of $z$ slightly greater than 2 and, therefore, 
are impractical for simulating large systems near a critical 
point. 

Cluster algorithms, on the other hand, such as the Swendsen-Wang 
(SW)~\cite{SwWa} algorithm, employ non-local update moves, flipping 
clusters of spins of linear extent comparable to the correlation length.  
This technique significantly reduces critical slowing down, and thus 
makes cluster algorithms preferable for simulating spin systems near a 
critical 
phase transition.  Recent numerical estimates of the SW dynamic exponent for 
two-dimensional ferromagnetic $q$-state Potts models are 
$z^{{\rm SW}}\approx .25$~\cite{BaCo} for the Ising ($q=2$) case and 
$z^{{\rm SW}}\approx .52$~\cite{SaSo97} for 
$q=3$.

Although there currently exists no theoretical means by which the 
dynamic exponent of a SW-type algorithm may be calculated, there is 
a rigorous lower bound.  The Li-Sokal~\cite{LiSo89,LiSo,ChMa} bound, 
as it has come to be known, states that $z^{{\rm SW}}\ge \alpha/\nu$ for 
$q$-state Potts models, where $\alpha$ and $\nu$ are the usual static 
critical exponents for the specific heat and correlation length, 
respectively.  We note that the numerical values given above are 
consistent with this bound, since, in two dimensions, $\alpha/\nu=0$ 
$(\log)$ for $q=2$, and $\alpha/\nu=2/5$ for $q=3$.  For the remainder 
of this paper we will continue to focus on the Ising and 3-state Potts 
models in two-dimensions, since these are the two most carefully 
studied cases.

We now turn to the invaded cluster 
(IC)~\cite{MaCh95,MaCh96,ChMa97,yscthesis} algorithm, a recent approach 
based on invasion percolation, for simulating equilibrium critical 
points.  This algorithm has the unique property that it ``self-organizes'' 
to the critical point.  Therefore, no {\em a priori} knowledge of the 
critical temperature is required; instead, $T_{{\rm c}}$ is an output of the 
algorithm.  In addition, due to an intrinsic negative-feedback 
mechanism, the IC algorithm equilibrates very quickly in 
the sense that thermodynamic quantities are measured to be near their 
equilibrium values within a few MC steps (after starting, say, from a 
completely ordered state).

Initial studies~\cite{MaCh96} seemed to indicate that the IC algorithm 
suffers no critical slowing down for the Ising model.  For both $d=2$ 
and $d=3$, the integrated autocorrelation time 
$\tau_{\varepsilon}$ was observed to decrease with $L$, 
while $\tau_{m}$ remained constant (within error bars), 
where $\varepsilon$ is the energy per spin, and $m$ is the fraction of 
spins in the largest cluster. (We omit the usual ``int'' subscript 
on $\tau$, since we deal almost exclusively with integrated as opposed 
to exponential autocorrelation times.) The decrease of 
$\tau_{\varepsilon}$ with $L$ was also observed for $3$- 
and $4$-state Potts models in two dimensions, but in these cases 
critical slowing was evident in the behavior of $\tau_{m}$.  
In Ref.~\cite{MaCh96} the dynamic exponents were estimated to be 
$z_{m}\approx .28$ for $q=3$ and $z_{m}\approx 
.63$ for $q=4$, each of which is less than the Li-Sokal bound on 
$z^{{\rm SW}}$ for its respective value of $q$.

In this paper, we repeat these studies for $q=2$ and $q=3$ in two 
dimensions using larger lattice sizes and an improved 
method~\cite{SaSo97} of estimating $\tau$.  We also investigate the 
$L$-dependence of the ``specific-heat-like'' quantity $c(L)\equiv 
L^{d}{\rm var}(\varepsilon)$.  (In the canonical ensemble $c$ is the 
specific heat, but the same is not true for the IC ensemble.)  In 
addition we measure $\tau_{\varepsilon}(l,L)$, the integrated 
autocorrelation time for the energy per spin $\varepsilon(l,L)$ 
measured on an $l\times l$ sub-lattice of the whole $L\times L$ 
lattice, as well as $c(l,L)\equiv l^{d}{\rm var}(\varepsilon(l,L))$.  
While the motivations for these experiments are discussed in more 
detail in \sect{crossover}, the central idea is that we expect the 
negative feedback to diminish for length scales $l<L$, leading to 
sub-system behavior that differs from that of the whole system.  We 
argue in \sect{crossover} that, for $l$ sufficiently less than $L$, we 
should observe a crossover to SW behavior for observables measured on 
a sub-lattice of size $l$.  Upon investigating the crossover region, 
we find that $c(l,L)$ has a simple scaling form and give an estimate 
of the length scale at which $c(l,L)$ crosses over to its SW analog, 
namely, the specific heat for an $l\times l$ system.

The results are less conclusive in the case of the dynamic variable 
$\tau_{\varepsilon}(l,L)$, but it appears that crossover to 
SW behavior does occur and that the crossover length for 
$\tau_{\varepsilon}(l,L)$ differs from that for $c(l,L)$ 
and is likely the same for $q=2$ and $q=3$.  In addition, the 
crossover phenomenon leads to a maximum in 
$\tau_{\varepsilon}(l,L)$ as $l$ is varied for a given $L$.  
We argue in \sect{crossover} that the dynamic exponent $z^{{\rm IC}}$ 
of the IC algorithm is appropriately defined by 
$\tau_{\varepsilon,{\rm max}}\sim L^{z^{{\rm IC}}}$.  For 
the 3-state Potts model we give a numerical estimate for $z^{{\rm IC}}$ 
that is smaller than a recent estimate~\cite{SaSo97} of $z^{{\rm SW}}$ and also 
less than the Li-Sokal bound on $z^{{\rm SW}}$.  For the Ising model the 
results are less conclusive, but it appears that $z^{{\rm IC}}<.21$
and possibly that $\tau_{\varepsilon,{\rm max}}\sim 
\log L$ so that $z^{{\rm IC}}=0$---similar to the state of affairs for 
the SW algorithm~\cite{BaCo}.

The remainder of this paper is organized as follows.  In \sect{ic} we 
provide some background on the invaded cluster algorithm and discuss 
results of previous IC simulations. In \sect{crossover} we discuss the 
crossover phenomenon in greater detail and propose a scaling form to 
relate observables measured on sub-lattices using the IC algorithm to 
corresponding quantities for the SW algorithm.  We describe our IC 
simulations of the Ising and 3-state Potts models in \sect{sim} and 
discuss the results in \sect{results}.  \sect{conclusion} contains our 
conclusions.

\section{Invaded cluster algorithm for critical Potts models}
\sectlabel{ic}

In order to understand how the IC algorithm works, we first review the 
SW algorithm for Potts models.  The ferromagnetic $q$-state Potts 
hamiltonian is
\begin{equation}
H=-\sum_{\langle i,j \rangle}\delta_{\sigma_{i},\sigma_{j}},
\end{equation}
where $\sigma_{i} \in \{0,1,\ldots,q-1\}$ and the sum is over 
nearest-neighbor spin pairs.  (Note that the Ising model is just the 
special case $q=2$.)

Given an initial spin configuration, the SW algorithm proceeds as 
follows:  First, satisfied bonds are occupied with probability 
$p=1-e^{-\beta}$, with $\beta=1/T$, where a bond joining two spins 
$i$ and $j$ is defined to be {\em satisfied} if and only if 
$\sigma_{i}=\sigma_{j}$.  Unsatisfied bonds are never occupied.  
Next, clusters of spins connected by occupied bonds are identified, 
and each cluster is ``flipped'', i.e., independently and uniformly 
assigned a new random spin value from $\{0,1,\ldots,q-1\}$.  (Note that 
a cluster can consist of a single spin.)  Finally, after statistics 
have been collected, occupied bonds are erased and the whole process 
is repeated.  It can easily be shown that the SW algorithm satisfies 
detailed balance for the canonical ensemble.

The IC algorithm uses invasion percolation to generate the spin 
clusters to be flipped.  Given an initial spin configuration, the 
first step is to assign a random order to the bonds of the lattice.  
The bonds are then examined, and satisfied bonds occupied one at a time 
in this order.  If a bond joining two clusters is occupied, they are 
combined into one.  Cluster growth continues until some stopping 
condition is fulfilled.  In this paper, we consider the topological 
spanning condition, which dictates that growth be stopped as soon as 
some cluster winds around the system in one of the $d$ directions.  As 
soon as spanning is detected, clusters (including the spanning 
cluster) are flipped exactly as in the SW algorithm, statistics are 
collected, bonds erased, and the process repeated.

To understand why the IC algorithm self-organizes to the critical 
point, we define $f$ to be the ratio of the number of occupied bonds 
to the number of satisfied bonds when some cluster first spans the 
system.  It has been argued~\cite{MaCh95,MaCh96,yscthesis} that, as 
the system size $L$ approaches infinity, the distribution of $f$ 
approaches a delta function at $p_{{\rm c}}\equiv 
1-e^{-\beta_{{\rm c}}}$, where $\beta_{{\rm c}}$ is the inverse critical 
temperature.  Though not a rigorous proof, the argument proceeds as 
follows.  First, we note that $p_{{\rm c}}$ is the 
threshold for percolation on the satisfied bonds of a critical spin 
configuration~\cite{ACCN}.  Thus, given a spin configuration that is typical of 
the critical point, the fraction $f$ of satisfied bonds that must be 
occupied to achieve spanning is close to $p_{{\rm c}}$.  Second, we 
observe that each iteration of the IC algorithm is identical to an 
iteration of the SW algorithm with $p=f$.  Therefore, performing an 
iteration of the IC algorithm on a critical spin configuration is 
equivalent to performing an iteration of the SW algorithm with 
$p \approx p_{{\rm c}}$, and thus the system will remain near the critical 
point.

If, instead, the system is started in the low-temperature phase, the 
number of satisfied bonds will be larger than is typical of 
$T_{{\rm c}}$.  Therefore, a smaller fraction $f$ will need to be 
occupied to achieve spanning.  In this case, an IC iteration is 
equivalent to a SW iteration with $p<p_{{\rm c}}$, i.e., 
$T>T_{{\rm c}}$, and therefore the system is pushed toward 
$T_{{\rm c}}$ from below.  Similarly, if the system is started in the 
high-temperature phase, it is pushed toward $T_{{\rm c}}$ from 
above.  Thus, in summary, a system in a non-critical state is pushed 
toward criticality, while a system in a critical state remains near 
criticality with $T$ fluctuating about $T_{{\rm c}}$.

Because of this negative-feedback mechanism, the IC algorithm 
self-organizes to the critical point with no {\em a priori} knowledge 
of $T_{{\rm c}}$.  Instead, $T_{{\rm c}}$ is obtained as an output of 
the algorithm, via the relation $\langle f \rangle=1-e^{-1/T_{{\rm c}}}$.  For 
example, results of IC simulations for the $2d$ Ising model yield an 
estimate of $T_{{\rm c}}\approx 1.1355$\cite{MaCh95} when extrapolated 
to $L=\infty$, as compared to the exact result for an infinite 
system, $T_{{\rm c}}= 1.1346\ldots$.

Since every configuration generated by the IC algorithm {\em must} 
contain a spanning cluster, it is clear that the algorithm does not 
sample the canonical ensemble for a system of finite volume.  We refer 
to the stationary distribution sampled by the IC algorithm as the IC 
ensemble.  If we assume that as $L\rightarrow\infty$ the distribution 
of $f$ approaches a delta function at $p_{{\rm c}}$ and that the volume 
fraction of the spanning cluster goes to zero, local observables such as 
internal energy and magnetization will approach their infinite-volume 
critical values in this limit.  Simulation 
results~\cite{MaCh95,MaCh96,yscthesis} support this hypothesis.  For 
example, results of IC simulations for the $2d$ Ising model yield an 
estimate for the energy per spin of $\varepsilon_{{\rm c}}\approx 
-1.706$~\cite{MaCh95} when extrapolated to $L=\infty$, as compared to 
the exact result for an infinite system, $\varepsilon_{{\rm c}}= 
-1.7071\ldots$.

Some finite-volume fluctuations in the IC ensemble, however, are very 
different from those in the canonical ensemble.  For example, in the 
canonical ensemble the quantity $c(L)\equiv 
L^{d}{\rm var}(\varepsilon)$ is the specific heat which diverges as 
$L^{\alpha/\nu}$ at the critical point---a logarithmic divergence for 
the Ising model in two dimensions.  In the IC ensemble, however, 
$c(L)$ is observed~\cite{MaCh95} to increase roughly linearly with $L$ 
for the $2d$ Ising model.  These differences can be traced to 
fluctuations in the effective temperature (measured by $f$) in the 
equilibrium state.  In the next section, we 
will examine more closely the roles played by temperature fluctuations 
and the negative-feedback 
mechanism in determining the properties of the IC algorithm.

\section{Crossover to Swendsen-Wang behavior}
\sectlabel{crossover}

As described in the previous section, the negative-feedback mechanism, 
inherent in the IC algorithm, drives the system to criticality by 
effectively adjusting the temperature after each iteration.  As 
previously mentioned, this mechanism leads to differences in the 
$L$-dependence of several dynamic and static quantities from that 
observed for the SW algorithm.  Now, however, we consider an $l\times 
l$ sub-lattice within the $L\times L$ lattice. Since the energy of a 
sub-system of size $l\ll L$ is weakly correlated with that of the 
whole system, the negative feedback mechanism is less effective for 
the sub-system.  A ``warm'' (relative to $T_{{\rm c}}$) sub-system in 
a ``cool'' system will be further warmed by the next IC iteration.  
As a result, the energy autocorrelation time for the sub-system may be 
longer than for the whole system.

The observation that an iteration of the IC algorithm is equivalent to 
an iteration of the SW algorithm with $p=f$ provides further insight 
into IC dynamics.  In particular, if the distribution of $f$ 
approaches a delta function as $L\rightarrow \infty$, then any finite 
sub-system of an infinite system will behave exactly as it would under 
SW dynamics. In short, for $l$ sufficiently smaller than $L$, the sub-system 
doesn't ``know'' it is being updated by the IC and not the SW 
algorithm.  Thus we expect that, in the limit $L\rightarrow \infty$, 
all static and dynamic quantities measured on a sub-system of finite size $l$ 
will approach the values measured for the SW algorithm for a 
sub-system of the same size.  It also follows that, for fixed $L$, there is a 
crossover from SW to IC behavior at intermediate values of $l$.  For 
example, the integrated autocorrelation time 
$\tau_{\varepsilon}(l,L)$ for the energy per spin in a 
sub-system of size $l$ should initially increase with $l$ as 
$l^{z^{{\rm SW}}}$ for $l\ll L$, reach a maximum, and then decrease as 
$l$ is increased further into the range where the negative-feedback 
mechanism becomes significant.

We can estimate the length scale at which the crossover occurs as 
follows.  In the canonical ensemble, the temperature uncertainty of 
the critical region scales 
with system size $L$ as $\delta T\sim L^{-1/\nu}$.  In the IC 
ensemble, however, temperature fluctuations are governed by the 
negative-feedback mechanism as described above.  In 
Ref.~\cite{yscthesis} it was found that the standard deviation of $f$ 
scales as $\sigma(f)\sim L^{-b}$ with 
$b\approx.46 (.30)$ for $q=2 (3)$ as compared with $1/\nu = 1 (6/5)$ 
respectively.  
Since the SW algorithm samples from the canonical ensemble and since 
we expect SW behavior for sub-systems of size $l\ll L$, crossover 
between the IC and SW regimes should occur when the temperature 
uncertainties from the two sources are comparable.  Thus, for a given 
$L$, we expect crossover at a sub-system size $l_{{\rm c}}$ given by 
$l_{{\rm c}}\sim L^{b\nu}$.  

Therefore, in light of the previous 
arguments, we hypothesize that the crossover from IC to SW behavior 
may be described by the scaling relationship
\begin{equation}
A^{{\rm IC}}(l,L)=A_{\infty}^{{\rm SW}}(l) F_{A}(l/L^{y}),
\eqnlabel{scalehyp}
\end{equation}
where $A_{\infty}^{{\rm SW}}(l)$ is any observable measured for the SW 
algorithm on an $l\times l$ sub-lattice immersed in an infinite 
system, $A^{{\rm IC}}(l,L)$ is the same observable measured for the IC 
algorithm on an $l\times l$ sub-lattice of an $L\times L$ lattice, 
$F_{A}$ is a scaling function with the property that 
$F_{A}(x)\rightarrow 1$ as $x\rightarrow 0$, and $y=b\nu$.  A word of 
caution concerning boundary conditions is in order here.  We define 
$A_{0}^{{\rm SW}}(L)$ to be an observable measured for the SW algorithm 
on an $L\times L$ lattice with periodic boundary conditions.  Although
we expect $A_{\infty}^{{\rm SW}}(l)/A_{0}^{{\rm SW}}(l)$ to approach a 
constant for $l\rightarrow \infty$, the constant will in general not 
be exactly 1 due to the different boundary conditions.

In this paper, we also seek to define a meaningful dynamic exponent 
$z^{{\rm IC}}$ for the IC algorithm that may be compared with exponents 
for other algorithms as well as with the Li-Sokal bound.  It is not 
obvious how to do this since the energy autocorrelation time 
$\tau_{\varepsilon}$ for the entire system was 
observed~\cite{MaCh96} to decrease with $L$.  However, this is not the whole 
story since we have argued above that correlations between 
successive IC configurations should decay more slowly on length scales 
$l<L$.  Since we would like $z^{{\rm IC}}$ to describe the 
$L$-dependence of the slowest mode, we suggest that it is the maximum 
value of $\tau_{\varepsilon}(l,L)$ attained for a given $L$ 
that is relevant.  Thus we define $z^{{\rm IC}}$ according to
\begin{equation}
\tau_{\varepsilon,{\rm max}}\sim L^{z^{{\rm IC}}}.
\eqnlabel{zdef}
\end{equation}
In the next section, we describe simulations designed to measure 
$z^{{\rm IC}}$ and to test the scaling hypothesis (\eqn{scalehyp}) for the static 
variable $c(l,L)$ as well as for the dynamic variable 
$\tau_{\varepsilon}(l,L)$.

\section{Description of simulations}
\sectlabel{sim}

We used the invaded cluster algorithm with the topological spanning 
rule to simulate the two-dimensional Ising and 3-state Potts models 
at their critical points for systems ranging in size from $L=32$ to 
$L=1024$.  Starting from a completely ordered state, we performed a 
number of relaxation steps to allow the system to reach equilibrium 
at $T_{{\rm c}}$ and then collected data for four observables: the energy 
per spin $\varepsilon$, the ratio $f$ of occupied to satisfied bonds, 
the fraction $m$ of spins in the largest cluster, and the 
susceptibility $\chi$, given by the sum of the squared cluster sizes 
divided by the total number of spins.  In addition to measuring the 
mean value and variance for each observable, we also measured the autocorrelation 
function and used this to calculate integrated autocorrelation 
times.  For a given observable $A$, the (normalized) 
autocorrelation function at a given time step $t$ can be calculated 
from a sequence of $n$ MC measurements $\{A(j), j=1,\ldots,n\}$ 
according to
\begin{equation}
\Gamma_{A}(t)\equiv\frac{\sum_{j=1}^{n-t}(A(j)-\langle A
\rangle)(A(j+t)-\langle A\rangle)}{\sum_{j=1}^{n}(A(j)-\langle A 
\rangle)^{2}},
\end{equation}
where $\langle A \rangle$ is the mean value of $A$.

The integrated autocorrelation time for the observable $A$ is defined 
by
\begin{equation}
\tau_{A}\equiv \frac{1}{2} + 
\sum_{t=1}^{\infty}\Gamma_{A}(t).
\end{equation}
Obviously, in practice, the sum must be truncated at some reasonable 
value of $t$.  Following the recommendation of Ref.~\cite{SaSo97}, we 
define
\begin{equation}
\tau_{A}(t_{A}^{*})\equiv \frac{1}{2} + 
\sum_{t=1}^{t_{A}^{*}}\Gamma_{A}(t)
\end{equation}
and choose the cutoff $t_{A}^{*}$ to be the smallest integer such 
that $t_{A}^{*}\ge\kappa\tau_{A}(t_{A}^{*})$, where 
$\kappa$ is a constant whose value will be discussed shortly.  If the 
autocorrelation function has the scaling form 
$\Gamma_{A}(t)=G(t/\tau_{{\rm exp}})$, where $\tau_{{\rm exp}}$ is the 
exponential autocorrelation time, then choosing the cutoff in this 
manner will insure that $\tau_{A}(t_{A}^{*})$ is 
proportional to $\tau_{A}$.  Thus estimates of 
$z_{A}$ will not be biased by truncating the sum at 
$t=t_{A}^{*}$.

One also would like the values of $\tau_{A}(t_{A}^{*})$ to 
approximate $\tau_{A}$ as accurately and precisely as 
possible, and here there is a tradeoff between excluding noise and 
including as much of the signal as possible.  In Ref.~\cite{SaSo97} 
it is shown that if $\Gamma_{A}(t)$ is roughly a single exponential, 
then choosing a value of $\kappa$ in the range 4--6 would achieve 
the optimal compromise for $n/\tau$ in the range $10^{4}$--$10^{6}$
that we used in our simulations.  However, although $\Gamma_{A}(t)$ 
is well approximated by a single exponential in the case of the SW 
algorithm, this is not true for the IC algorithm~\cite{MaCh96}.  For 
this reason, and since we are willing to accept slightly larger 
statistical uncertainties in order to reduce systematic errors, we 
used $\kappa=10$ in all our calculations.

In light of the discussion in \sect{crossover}, we also collected 
data for several sub-system sizes for each $L$.  Here we concentrated 
on the energy per spin, measuring $\langle \varepsilon(l,L) 
\rangle$, ${\rm var} (\varepsilon (l,L))$, and 
$\tau_{\varepsilon}(l,L)$ for sub-system sizes ranging from 
$l=1$ to $l=L/2$ for each $L$.  The sub-systems are squares all 
sharing a single corner of the lattice.   

Error bars on all quantities were calculated using the blocking 
method.  Each run was partitioned into $k$ contiguous blocks 
of $n$ MC steps each, and the individual blocks treated as independent 
runs.  Although this is an approximation, it will be a good one 
provided that $n$ is large compared to the system's longest 
relaxation time.  As an example of the blocking method, we obtain the 
value for, say, $\tau_{m}$ by first calculating 
$\tau_{m}^{(i)}$ for each block $i$.  We then calculate the 
mean
\begin{equation}
\overline{\tau_{m}}\equiv
\frac{1}{k}\sum_{i=1}^{k}\tau_{m}^{(i)}
\end{equation}
and its standard error
\begin{equation}
\sigma(\overline{\tau_{m}})\equiv
\sqrt{\frac{\sum_{i=1}^{k}(\tau_{m}^{(i)} -
\overline{\tau_{m}})^{2}}{k(k-1)}}
\end{equation}
and report the result 
$\tau_{m}=\overline{\tau_{m}}
\pm\sigma(\overline{\tau_{m}})$.

For both $q=2$ and $q=3$, one long run was initially performed for 
each lattice size $L$, and the blocking method implemented as just 
described.  (Three independent runs were performed for the case $q=2$, 
$L=1024$.)  The number of blocks used ranged from 100 for $L=32$ down 
to 10 for $L=1024$, and the number $n$ of MC steps per block ranged 
from $5\times 10^{3}$ to $1\times 10^{5}$.  In each case, $n$ was 
greater than the longest observed autocorrelation time for the given 
system by at least a factor of $10^{3}$ ($10^{4}$ for the smaller 
lattices), thereby making the assumption of independent blocks, used 
in calculating the error bars, a reasonably good one.  The number of 
equilibration steps performed at the beginning of each run also 
exceeded the longest observed $\tau$ by a factor of $10^{3}$ in all 
cases.

In these initial runs, we collected data for sub-system sizes 
$l\in\{1,2,4,\ldots,L/2\}$ as well as for the whole system ($l=L$).  
In order to estimate $z^{{\rm IC}}$ as defined in \eqn{zdef}, we sought 
to obtain an accurate value for 
$\tau_{\varepsilon,{\rm max}}(L)$ for each $L$.  Therefore, 
once we had learned, from the initial runs, the approximate sub-system 
size $l_{{\rm max}}$ at which $\tau_{\varepsilon}(l,L)$ 
attains a maximum, we then performed between one and three additional 
independent runs for each system and collected data for evenly spaced 
values of $l$ near our rough estimate of $l_{{\rm max}}$.  The entire 
experiment required about 5 months of CPU time on a single processor 
of a dual-processor 266MHz Pentium~II Linux workstation.  We used the 
machine-supplied random number generator coupled with a shuffling 
procedure as described in Ref.~\cite{yscthesis}.

\section{Discussion of results}
\sectlabel{results}

The first quantity we consider is the static variable $c(L)\equiv 
L^{d}{\rm var}(\varepsilon)$ (see \tab{csigf}).  In the canonical 
ensemble, $c(L)$ is the specific heat which diverges, at the critical 
point, as $\log L$ for $q=2$ and as $L^{\alpha/\nu}$ with 
$\alpha/\nu=2/5$ for $q=3$.  We see from \fig{cvsL}, however, that the 
situation is quite different for the IC ensemble, as first observed in 
Ref.~\cite{MaCh95}.  We assume that the asymptotic behavior is given 
by a power law $c(L)\sim L^{w}$ and fit a line to a plot of $\log_{10} 
c(L)$ vs $\log_{10} L$ as shown in \fig{cvsL}.  Note that where error 
bars are not visible they are smaller than the symbol height.  As is 
always the case when trying to ascertain asymptotic behavior from 
simulations at finite $L$, there can be some debate as to which, if 
any, data points should be omitted from the fit because of corrections 
to scaling.  Here and elsewhere in our analysis, we proceed by 
dropping points one at a time in order of increasing L until either 1) 
a reasonably good fit is obtained, 2) the fit ceases to improve 
significantly with further cuts, or 3) we are left with only three 
data points.  We employ standard, weighted $\chi^{2}$ fitting, using 
the confidence level~\footnote{The confidence level is the probability 
that a $\chi^{2}$ as poor as the measured value would occur, assuming 
that the underlying model is correct and that the measurement errors 
are normally distributed~\cite{PrTeVeFl}.  (Note that confidence level 
is denoted by the symbol $Q$ in Ref.~\cite{PrTeVeFl}.)} (CL) as our 
goodness-of-fit measure, and consider a fit to be ``reasonably good'' 
if ${\rm CL}\ge 10\%$.

For the Ising model, a fit to the last four data points ($L\ge 128$) 
yields $w=1.020 \pm .003$ (${\rm CL}=16\%$) in agreement with the 
observation $w\approx 1$ reported in Ref.~\cite{MaCh95}.  In the case 
of the 3-state Potts model, a fit to the last three points gives 
$w=1.313\pm .008$, but, because of the poor confidence level (2\%) and 
the upward curvature visible in the data, this value should probably 
just be regarded as a lower bound on $w$ for $q=3$.  We emphasize that 
the error bars on these and subsequent exponent estimates are purely 
statistical in nature and do not reflect the uncertainty of 
extrapolating to infinite system size.

Turning now to the dynamic variables (see \tab{taus}), we plot the 
logarithms of the 
autocorrelation times $\tau_{\varepsilon}$, $\tau_{f}$, $\tau_{m}$, 
and $\tau_{\chi}$ vs.\ $\log_{10} L$ for $q=2$ in \fig{q2taus} and for $q=3$ in 
\fig{q3taus}.  For $q=2$ we find that $\tau_{\varepsilon}$ and 
$\tau_{f}$ decrease with $L$ (perhaps in a rather complicated fashion) 
and $\tau_{m}$ and $\tau_{\chi}$ remain approximately constant for the 
range of $L$ values used in our simulations.  These results are in 
agreement with initial observations~\cite{MaCh95} that led to 
speculation of no critical slowing.  For $q=3$, however, we observe 
critical slowing in the behavior of $\tau_{m}$ and $\tau_{\chi}$.  
Fits to the data for $L\ge 64$ yield $z_{m}=.191\pm .004$ 
(${\rm CL}=25\%$) and
$z_{\chi}=.206\pm .005$ (${\rm CL}=20\%$), but it seems likely that 
$\tau_{m}\sim \tau_{\chi}$ in the asymptotic limit.  We note that the 
value of $z_{m}$ is somewhat smaller than the previous estimate 
$z_{m}\approx .28$~\cite{MaCh96}.

Next we consider $\tau_{\varepsilon}(l,L)$, the integrated 
autocorrelation time for the energy per spin $\varepsilon(l,L)$ 
measured on an $l\times l$ sub-lattice of the whole $L\times L$ 
lattice.  As expected from the discussion in \sect{crossover}, we see 
that as $l$ is increased for a given $L$, 
$\tau_{\varepsilon}(l,L)$ increases, reaches a maximum, and 
then decreases, as shown in \fig{q2tauvsl} for $q=2$ and in 
\fig{q3tauvsl} for $q=3$.

To find $\tau_{\varepsilon,{\rm max}}$ and its location $l_{{\rm max}}$ 
for a given $L$ (see \tab{lmaxtaumax}), we fit a parabola to the 
region of the curve near the maximum.  In order to do this 
objectively, we began, for each $L$, by omitting the data point with 
the smallest value of $\tau_{\varepsilon}$ and performing the fit.  We 
then dropped the point with the next smallest $\tau_{\varepsilon}$, 
re-fit, and continued in this fashion until 1) a CL of 50\% or greater 
was obtained and 2) the values of
$\tau_{\varepsilon,{\rm max}}$ and $l_{{\rm max}}$ that were 
obtained by dropping an additional point remained within error bars 
of the current best-fit values.

We then attempted to determine the dynamic exponent $z^{{\rm IC}}$, 
defined in \eqn{zdef}, by fitting a line to a plot of $\log_{10} 
\tau_{\varepsilon,{\rm max}}(L)$ vs.\ $\log_{10} L$ for $q=2$ and 
$q=3$ .  The results are shown in \fig{taumaxvsL} along with results 
for $\tau_{\varepsilon}^{{\rm SW}}$ taken from Baillie and 
Coddington~\cite{BaCo} for $q=2$ and from Salas and 
Sokal~\cite{SaSo97} for $q=3$.  We note that the increase of 
$\tau_{\varepsilon,{\rm max}}(L)$ with $L$ for $q=2$ is the 
first observation of critical slowing for the IC algorithm in 
the case of the Ising model.

For $q=2$ the autocorrelation times for the IC algorithm are nearly the same as for
the Wolff algorithm and smaller than those for the SW algorithm by a factor of about
2 (see \tab{lmaxtaumax}), but the  $L$-dependence is similarly obscure.  A good fit to a power law could 
not be obtained for the IC data for $q=2$.  The line shown in the figure, 
having slope $\approx .21$ is the best fit for $L\ge 64$, but it 
clearly does not describe the data very well.  The best power-law fit to the SW 
data for $64 \le L \le 512$ (also shown in the figure) yields 
$z^{{\rm IC}}\approx .25$ as reported in Ref.~\cite{BaCo}.  Although 
the fit is considerably better than that for the IC data, the CL is 
still poor ($<0.1\%$), and the better fit might be primarily due to the 
absence of data for $L=1024$ in the SW case.

Since it has been suggested~\cite{HeBu} that 
$\tau_{\varepsilon}^{{\rm SW}}$ increases logarithmically with $L$ 
rather than as a power of $L$, Baillie and Coddington also fit their 
data to a logarithm with somewhat better results (${\rm CL}=13\%$).  
For the IC algorithm, a logarithmic fit is still atrocious, albeit 
somewhat better than the power law.  Later in this section we 
consider the possibility that we have underestimated the error bars on 
$\tau_{\varepsilon,{\rm max}}$, which, of course, could 
result in a poor fit even if the underlying model had been correctly 
identified.  Still, even the general trend in the data is difficult 
to discern, indicating that corrections to scaling are probably significant for 
the system sizes studied here.  Therefore, we conclude that high 
precision data for larger lattices are needed before a more 
definitive statement can be made concerning the asymptotic behavior of 
$\tau_{\varepsilon,{\rm max}}$.

For the 3-state model, however, the picture appears to be somewhat 
clearer.  Although a slight downward curvature in the data is visible 
in \fig{taumaxvsL}, a good fit (${\rm CL}=69\%$) to a power law is 
obtained for $256\le L \le 1024$, yielding $z^{{\rm IC}}=.346\pm 
.002$.  This is to be compared with the value of 
$z^{{\rm SW}}_{\varepsilon}=.515 \pm .006$ obtained by Salas and Sokal 
for the SW algorithm with $128\le L\le 1024$ (${\rm CL}=80\%$).  For 
the Wolff algorithm $z^{{\rm Wolff}}_{\varepsilon}=.57 \pm .01$ was 
reported in Ref.~\cite{BaCo}.   While 
there is no guarantee that $\tau_{\varepsilon,{\rm max}}$ is  the system's longest
relaxation time, it is interesting that  $z^{{\rm IC}}$ is significantly smaller
than  $z^{{\rm SW}}_{\varepsilon}$ and also less than the Li-Sokal bound 
($z^{{\rm SW}}\ge \alpha/\nu = 2/5$).

Now we proceed to test the scaling hypothesis (\eqn{scalehyp}) 
presented in \sect{crossover}.  In order to do this, we first need to 
determine the exponent $b$ defined by  $\sigma(f)\sim L^{-b}$.  We 
plot $\log_{10} \sigma(f)$ vs.\ $\log_{10} L$ in \fig{sigfvsL} (the 
data are listed in \tab{csigf}) for $q=2$ and 
$q=3$ along with the best-fit lines.  A fit to all six $q=2$ points 
yields $b=.4781 \pm .0006$ (${\rm CL}=54\%$), while, for $q=3$ a good 
fit (${\rm CL}=68\%$) is obtiained for the last four points ($L\ge 
128$), resulting in $b=.3252\pm .0009$.  These results are consistent 
with previous estimates~\cite{MaCh96}. 

We now test \eqn{scalehyp} for the variable $c^{{\rm IC}}(l,L)\equiv 
l^{d}{\rm var}(\varepsilon(l,L))$.  Since $c_{0}^{{\rm SW}}(l)\sim \log 
l$ and we expect $c_{\infty}^{{\rm SW}}(l)\sim c_{0}^{{\rm SW}}(l)$ 
(recall $c_{0}^{{\rm SW}}(l)$ is the specific heat for an $l\times l$ 
lattice with periodic boundary conditions, and 
$c_{\infty}^{{\rm SW}}(l)$ is the specific heat for an $l\times l$ 
sub-lattice immersed in an infinite system), we plot 
$c^{{\rm IC}}(l,L)/\log_{10} l$ vs.\ $l/L^{y}$ in \fig{q2ccollapse}, 
where $y=b=.4781$ for the Ising model ($\nu = 1$).  The observed data 
collapse provides strong support for \eqn{scalehyp}.  For $q=3$, 
however, it was found in Ref.~\cite{SaSo97} that the asymptotic form 
$c_{0}^{{\rm SW}}(l)\sim l^{\alpha/\nu}$ does not describe the SW data 
very well for the range of lattice sizes considered here.  Therefore, 
we cannot expect \eqn{scalehyp} to provide a good description of the 
IC data if the asymptotic form is used for $c_{\infty}^{{\rm SW}}(l)$.  
Nevertheless, if we plot $c^{{\rm IC}}(l,L)/c_{0}^{{\rm SW}}(l)$ vs.\ 
$l/L^{y}$, using the measured values of $c_{0}^{{\rm SW}}(l)$ from 
Ref.~\cite{SaSo97} and $y=b\nu=.2711$ for $q=3$, data collapse is 
apparent in \fig{q3ccollapse}, although perhaps a bit less convincing 
than for the Ising case.

We note that the curve in \fig{q3ccollapse} extrapolates to about 
$.85$ on the vertical axis for $l/L^{y}=0$.  As previously mentioned, 
we would expect this value to be 1 if SW data were collected for 
sub-systems immersed in larger systems so as to reproduce the boundary 
conditions applied to the IC sub-systems.  Thus we conclude that our 
scaling hypothesis (\eqn{scalehyp}) does appear to be valid for the 
variable $c^{{\rm IC}}(l,L)$ for $q=2$ and $q=3$.

Although the static quantity $c^{{\rm IC}}(l,L)$ seems to be well 
described by \eqn{scalehyp} with $y=b\nu$, the same is not true for 
the dynamic quantity $\tau_{\varepsilon}(l,L)$.  This is 
easy to see, since \eqn{scalehyp} predicts that the location 
$l_{{\rm max}}$ of the maximum in Figs.~\ref{fig:q2tauvsl} 
and~\ref{fig:q3tauvsl} should scale as $L^{y}$; however, the plots of 
$\log_{10} l_{{\rm max}}$ vs.\ $\log_{10} L$ shown in \fig{lmaxvsL} reveal that 
this is not the case---at least not if $y=b\nu$ is required.  
For both $q=2$ and $q=3$ the slope of the 
best-fit line for $64\le L\le 1024$, shown in \fig{lmaxvsL}, 
is approximately $.62$, although the 
confidence levels are poor.  Unlike the situation encountered earlier 
in this section, when fitting $\tau_{\varepsilon,{\rm max}}$ 
to a power law in $L$, the points seem to be scattered randomly about 
the best-fit line.  Therefore, we suspect that $l_{{\rm max}}$ does 
scale as a power of $L$, but that our error bars on $l_{{\rm max}}$ are 
somewhat underestimated.

There are three aspects of our analysis that could lead to 
underestimates in the error bars on  
$\tau_{\varepsilon,{\rm max}}$ and $l_{{\rm max}}$. First, there always 
exists the possibility that the assumption of normally distributed 
measurement errors is not valid.  Second, 
the blocking method, used to calculate error bars on values of 
$\tau_{\varepsilon}(l,L)$, treats successive blocks as if 
they were independent runs, an approximation that may not be entirely 
justified even though the block length was greater than 
$10^{3}\tau_{\varepsilon,{\rm max}}$ in all cases.  Finally, 
and probably most importantly, the error bars on 
$\tau_{\varepsilon,{\rm max}}$ and $l_{{\rm max}}$ resulting 
from the weighted $\chi^{2}$ fit to a parabola are calculated by 
assuming that the measurements of 
$\tau_{\varepsilon}(l,L)$ at different values of $l$ for a 
given $L$ are independent.  This is clearly not a good approximation, 
since all of the sub-lattices extend outward from the same corner of 
the $L\times L$ lattice.  Therefore, all the spins in a given 
$l\times l$ sub-system are also contained in every larger sub-system, 
and thus sub-systems for comparable values of $l$ are highly 
correlated.

In any case, it is still clear that \eqn{scalehyp} with $y=b\nu$ 
cannot explain the data for $\tau_{\varepsilon}$.  
Since the logic leading to the scaling form seems sound, we 
hypothesize that \eqn{scalehyp} does hold for 
$\tau_{\varepsilon}$ but that the crossover length for
$\tau_{\varepsilon}$ is different from that for $c$ so 
that $y\ne b\nu$ in the case of $\tau_{\varepsilon}$.  This 
seems plausible, since there is no {\em a priori} reason why the 
thermodynamic argument by which we arrived at $y=b\nu$ must apply to 
the dynamic quantity $\tau_{\varepsilon}$.  Nevertheless, if
\eqn{scalehyp} still holds for $\tau_{\varepsilon}$, we can 
obtain the crossover exponent $y$ from \fig{lmaxvsL} as described above.

To test our scaling hypothesis for $\tau_{\varepsilon}$, we plot 
$\tau_{\varepsilon}(l,L)/\tau_{\varepsilon,0}^{{\rm SW}}(l)$
vs.\ $l/L^{y}$ with $y=.6176$ ($y=.626$) for $q=2$ ($q=3$) in 
\fig{q2taucollapse} (\fig{q3taucollapse}).  The values of 
$\tau_{\varepsilon,0}^{{\rm SW}}(l)$ are taken from Ref.~\cite{BaCo} for 
$q=2$ and from Ref.~\cite{SaSo97} for $q=3$.  The data collapse is 
not terribly convincing in either case, but seems too good to 
completely rule 
out \eqn{scalehyp} as the correct asymptotic form.  The fact that 
both curves extrapolate to about 1 for  $l/L^{y}=0$ provides further 
support for the scaling hypothesis.  Still, it appears that additional tests 
are needed to confirm or disprove \eqn{scalehyp} for 
$\tau_{\varepsilon}$.

\section{Conclusion}
\sectlabel{conclusion}

Using the invaded cluster (IC) algorithm with the topological spanning 
rule, we simulated the critical Ising and 3-state Potts models in two 
dimensions for systems ranging in size from $L=32$ to $L=1024$.  In 
accord with previous results~\cite{MaCh95,MaCh96}, we find that the 
$L$-dependence of several static and dynamic quantities is very 
different from that observed for the Swendsen-Wang (SW) algorithm 
which samples from the canonical ensemble.  In particular, the 
quantity $c(L)\equiv L^{d}{\rm var}(\varepsilon)$ is not proportional 
to the specific heat and the integrated autocorrelation time 
$\tau_{\varepsilon}$ for the energy per spin decreases with $L$.  
However, we find that the corresponding quantities 
$\tau_{\varepsilon}(l,L)$ and $c(l,L)$, measured for a sub-system of 
size $l$, exhibit a crossover to SW behavior for $l$ sufficiently 
less than $L$.  

To describe the crossover phenomenon, we propose the scaling form 
$A^{{\rm IC}}(l,L)=A_{\infty}^{{\rm SW}}(l) F_{A}(l/L^{y})$, where 
$A_{\infty}^{{\rm SW}}(l)$ is any observable measured for the SW 
algorithm on an $l\times l$ sub-lattice immersed in an infinite 
system, $A^{{\rm IC}}(l,L)$ is the same observable measured for the IC 
algorithm on an $l\times l$ sub-lattice of an $L\times L$ lattice, and 
$F_{A}$ is a scaling function with the property that 
$F_{A}(x)\rightarrow 1$ as $x\rightarrow 0$.  We have argued that the 
crossover exponent $y$ should equal $b\nu$, where $\nu$ is the usual 
correlation-length exponent and $b$ is defined by $\sigma(f)\sim 
L^{-b}$ with $f$ the ratio of occupied to satisfied bonds.  We find 
that the proposed scaling form with $y=b\nu$ provides a good 
description of our data for the static variable $c(l,L)$, but is less 
successful for the dynamic variable $\tau_{\varepsilon}(l,L)$, even if 
the possibility $y\ne b\nu$ is admitted.

In addition, we define the dynamic exponent $z^{{\rm IC}}$ for the 
invaded cluster algorithm in terms of the maximum 
value $\tau_{\varepsilon,{\rm max}}$ attained for a given $L$ according 
to $\tau_{\varepsilon,{\rm max}}\sim L^{z^{{\rm IC}}}$.  For $q=3$ we 
find that  $z^{{\rm IC}}=.346\pm .002$, which is smaller 
than recent numerical estimates of $z^{{\rm SW}}$ and $z^{{\rm Wolff}}$ 
and also less 
than the Li-Sokal bound on $z^{{\rm SW}}$.  For $q=2$ we also observe 
critical slowing, but the $L$-dependence of $\tau_{\varepsilon,{\rm max}}$ 
is less clear.  It appears from our simulations that $z^{{\rm IC}}<.21$ and 
possibly that $\tau_{\varepsilon,{\rm max}}\sim \log L$  
($z^{{\rm IC}}=0$), but  high 
precision data for larger lattices are needed before a more 
definitive statement can be made concerning the asymptotic behavior of 
$\tau_{\varepsilon,{\rm max}}$ for the Ising model.

\section*{Acknowledgements}
We are grateful to Yongsoo Choi for providing the original code for 
the invaded cluster algorithm and to Robert Guyer for useful 
discussions concerning the analysis.  This work was supported in part 
by NSF Grant DMR-9632898 and NSA Grant MDA 904-98-1-0518.

\newpage

\begin{table}  
  \caption{$c\equiv L^{d}{\rm var}(\varepsilon)$ and $\sigma(f)$
  for the IC algorithm for the $2d$ Ising and 3-state Potts models, 
  where $\varepsilon$ is the energy per spin, $f$ is the ratio of occupied 
  to satisfied bonds, and $L$ is the lattice size.}
 \tablabel{csigf}  
 \begin{tabular}{rllll} 
$L$&$c\ (q=2)$&$c\ (q=3)$&$\sigma(f)\ \ (q=2)$&$\sigma(f)\ \ (q=3)$\\
\tableline 
   32 &  3.288(3)       &  5.067(4)    & \ \ .05100(4) & \ \ .06991(4) \\
   64 &  6.038(6)       &  11.239(7)   & \ \ .03657(3) & \ \ .05508(2) \\
  128 &  11.87(2)       &  26.24(2)    & \ \ .02626(3) & \ \ .04367(3) \\
  256 &  24.03(8)       &  63.0(1)     & \ \ .01886(4) & \ \ .03482(4) \\
  512 &  48.4(3)        &  155.0(6)    & \ \ .01352(7) & \ \ .02781(9) \\
 1024 &  99.7(6)        &  390(2)      & \ \ .00977(4) & \ \ .0223(1)  \\
 \end{tabular}
\end{table}

\begin{table}  
  \caption{Integrated autocorrelation times for the IC algorithm for 
  the $2d$ Ising and 3-state Potts models.
  $\varepsilon$ is the energy per spin, $f$ is the ratio of occupied to 
  satisfied bonds, $m$ is the fraction of spins in the largest cluster, 
  $\chi$ is the susceptibility, and $L$ is the lattice size.}
 \tablabel{taus}  
 \begin{tabular}{crllll} 
$q$&$L$&$\ \ \ \tau_{\varepsilon}$&$\ \ \ \ \tau_{f}$&
$\ \ \ \tau_{m}$&$\ \ \ \tau_{\chi}$\\
\tableline 
2 &   32 & .546(1)             & .1828(7) & .857(3)    & .798(2) \\
2 &   64 & .499(2)             & .1275(7) & .852(3)    & .795(3) \\
2 &  128 & .443(2)             & .0845(7) & .859(3)    & .802(3) \\
2 &  256 & .384(3)             & .070(3)  & .869(8)    & .807(7) \\
2 &  512 & .305(3)             & .065(3)  & .88(1)     & .82(1)  \\
2 & 1024 & .260(3)             & .033(2)  & .90(1)     & .83(1)  \\ 
\\
3 &   32 & .832(2)             & .1835(5) & 1.303(4)    & 1.208(4) \\
3 &   64 & .821(2)             & .1742(4) & 1.435(4)    & 1.351(4) \\
3 &  128 & .753(2)             & .1369(5) & 1.629(7)    & 1.552(7) \\
3 &  256 & .635(3)             & .1071(8) & 1.89(1)     & 1.81(1)  \\
3 &  512 & .516(5)             & .079(1)  & 2.13(3)     & 2.08(3)  \\
3 & 1024 & .407(4)             & .053(1)  & 2.38(5)     & 2.32(5)  \\
 \end{tabular}
\end{table}

\begin{table}  
  \caption{$l_{{\rm max}}$ and autocorrelation times for the $2d$ Ising and 3-state 
  Potts models.  $L$ is the lattice size, $l_{{\rm max}}$ and 
  $\tau_{\varepsilon,{\rm max}}^{{\rm IC}}$ are the location and height, 
  respectively, of the maximum in Figs.~\ref{fig:q2tauvsl} 
  and~\ref{fig:q3tauvsl}.  $\tau_{\varepsilon}^{{\rm SW}}$ and 
  $\tau_{\varepsilon}^{{\rm Wolff}}$ are the 
  integrated energy autocorrelation times for the SW and 
  Wolff algorithms.}
 \tablabel{lmaxtaumax}  
 \begin{tabular}{crllll} 
$q$&$L$&$\ \ l_{{\rm max}}$&$\ \tau_{\varepsilon,
{\rm max}}^{{\rm IC}}$&
$\ \tau_{\varepsilon}^{{\rm SW}}$\tablenotemark[1]&$\ \tau_{\varepsilon}^{{\rm 
   Wolff}}$\tablenotemark[2]\\
\tableline 
2 &   32 &  8.36(5)  &  1.962(3)   & 4.016(5) & 1.815(3) \\
2 &   64 &  13.89(8) &  2.319(2)   & 4.90(1)  & 2.225(6) \\
2 &  128 &  21.1(4)  &  2.694(7)   & 5.87(2)  & 2.654(12)\\
2 &  256 &  32.2(3)  &  3.133(6)   & 6.87(3)  & 3.076(24)\\
2 &  512 &  51.2(6)  &  3.60(1)    & 8.0(1)   & --       \\
2 & 1024 &  75.9(9)  &  3.90(2)    & --       & --       \\ 
\\
3 &   32 &  9.53(7)  & 4.206(8)  &  13.28(6)  & 8.76(4)  \\
3 &   64 &  15.9(2)  & 5.55(1)   &  19.5(1)   & 13.08(16)\\
3 &  128 &  25.1(2)  & 7.17(1)   &   28.5(1)  & 19.5(3)  \\
3 &  256 &  37.0(3)  & 9.003(7)  &   40.8(2)  & 27.7(8)  \\
3 &  512 &  63.2(9)  & 11.46(5)  &   58.5(6)  & --       \\
3 & 1024 &  90(1)    & 14.5(1)   &   82.2(2)  & --       \\
 \end{tabular}\tablenotetext[1]{From Ref.~\cite{BaCo} for $q=2$ and 
 Ref.~\cite{SaSo97} for $q=3$.}\tablenotetext[2]{From Ref.~\cite{BaCo}.}
\end{table}  

\newpage

\begin{figure}
 \begin{center}
  \begin{picture} (274, 462) (0,-63) 
   \put(0,0){\epsfbox[140 289 414 563]{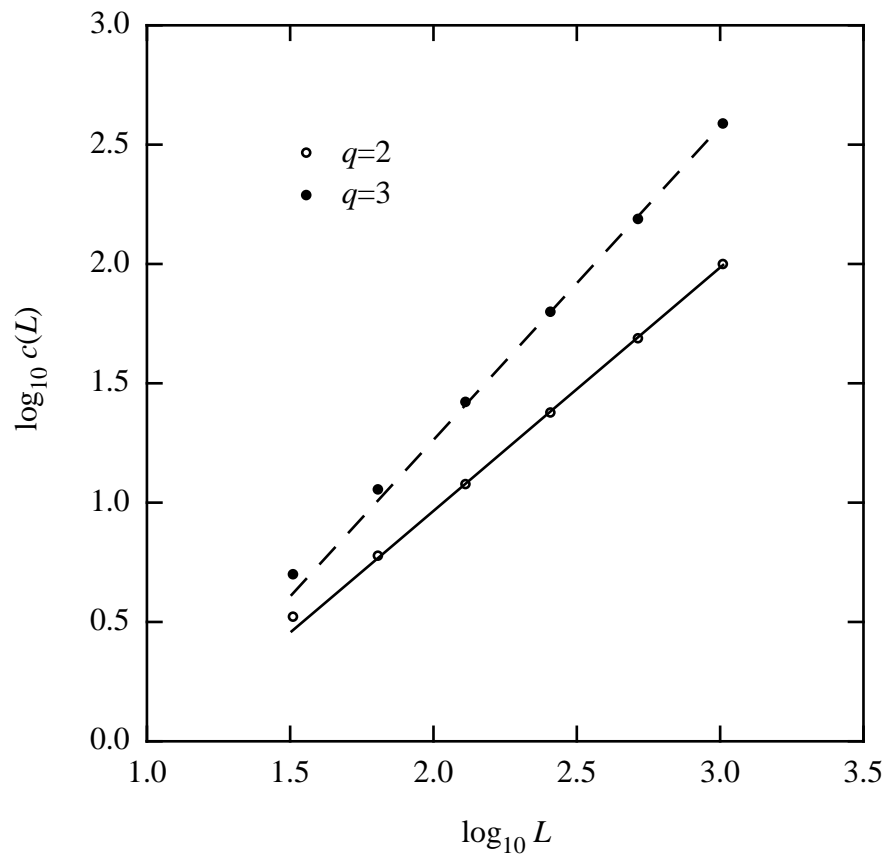}}
  \end{picture}
  \caption{$\log_{10} c(L)$ vs.\ $\log_{10} L$ for the IC algorithm, plotted 
  for the $2d$ Ising and 3-state Potts models.  Here, $c(L)\equiv 
  L^{d}{\rm var}(\varepsilon)$, where $\varepsilon$ is the energy per 
  spin and $L$ is the lattice size.  The solid (dashed) line is a 
  linear fit to the $q=2$ ($q=3$) data for $128\le L\le 1024$ ($256\le 
  L\le 1024$) and has slope 1.020 (1.313).} 
  \figlabel{cvsL}
 \end{center}
\end{figure}

\newpage

\begin{figure}
 \begin{center}
  \begin{picture} (274, 462) (0,-63)
   \put(0,0){\epsfbox[140 289 414 563]{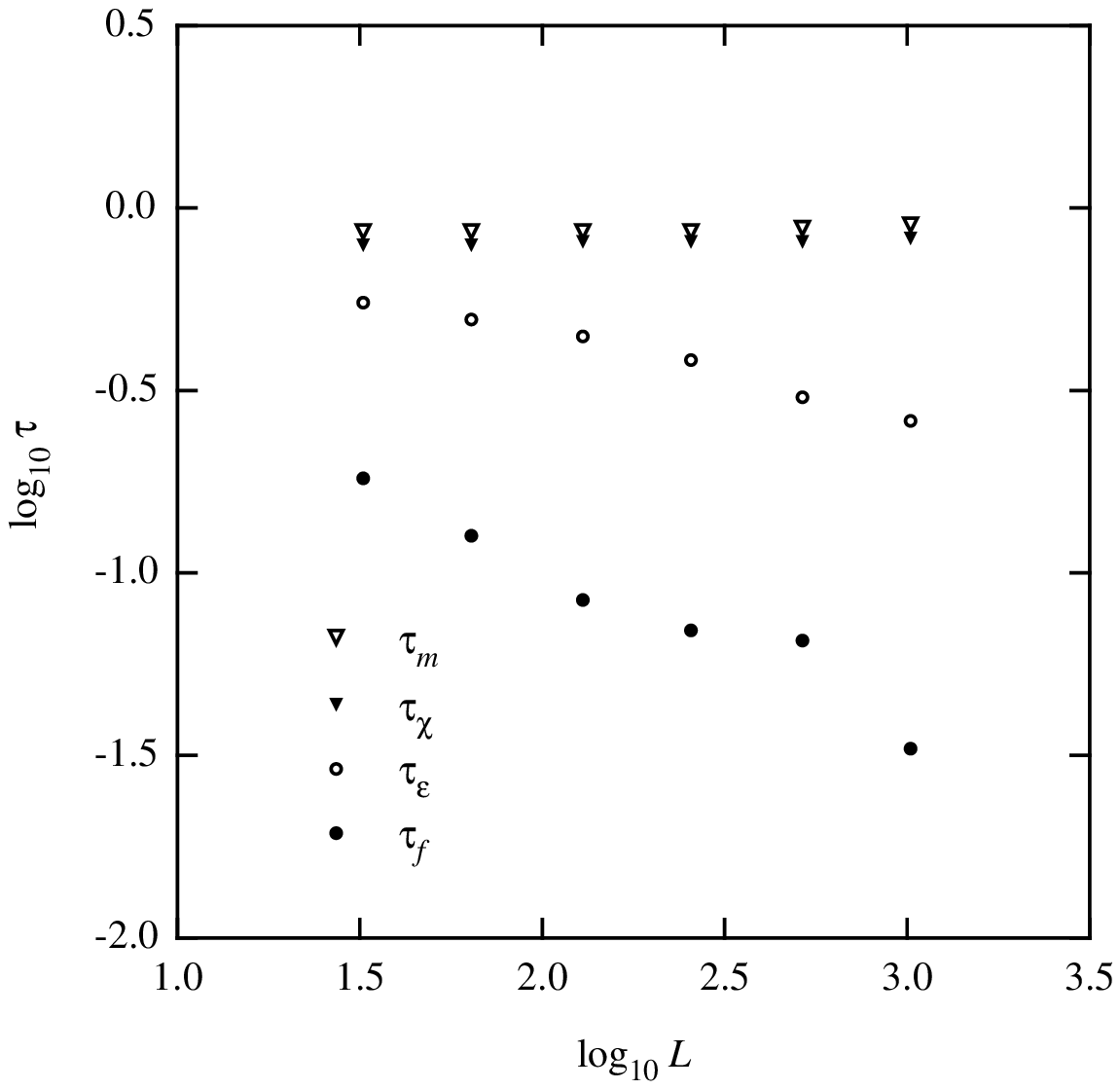}}
  \end{picture}
  \caption{$\log_{10} \tau$ vs.\ $\log_{10} L$ for the IC algorithm, where 
  $\tau$ is the integrated autocorrelation time and $32\le L \le 1024$ 
  is the lattice size, plotted for the $2d$ Ising model for the 
  energy per spin $\varepsilon$, the ratio $f$ of occupied to 
  satisfied bonds, the fraction $m$ of spins in the largest cluster, 
  and the susceptibility $\chi$.} 
  \figlabel{q2taus}
 \end{center}
\end{figure}

\newpage

\begin{figure}
 \begin{center}
  \begin{picture} (274, 462) (0,-63) 
   \put(0,0){\epsfbox[140 289 414 563]{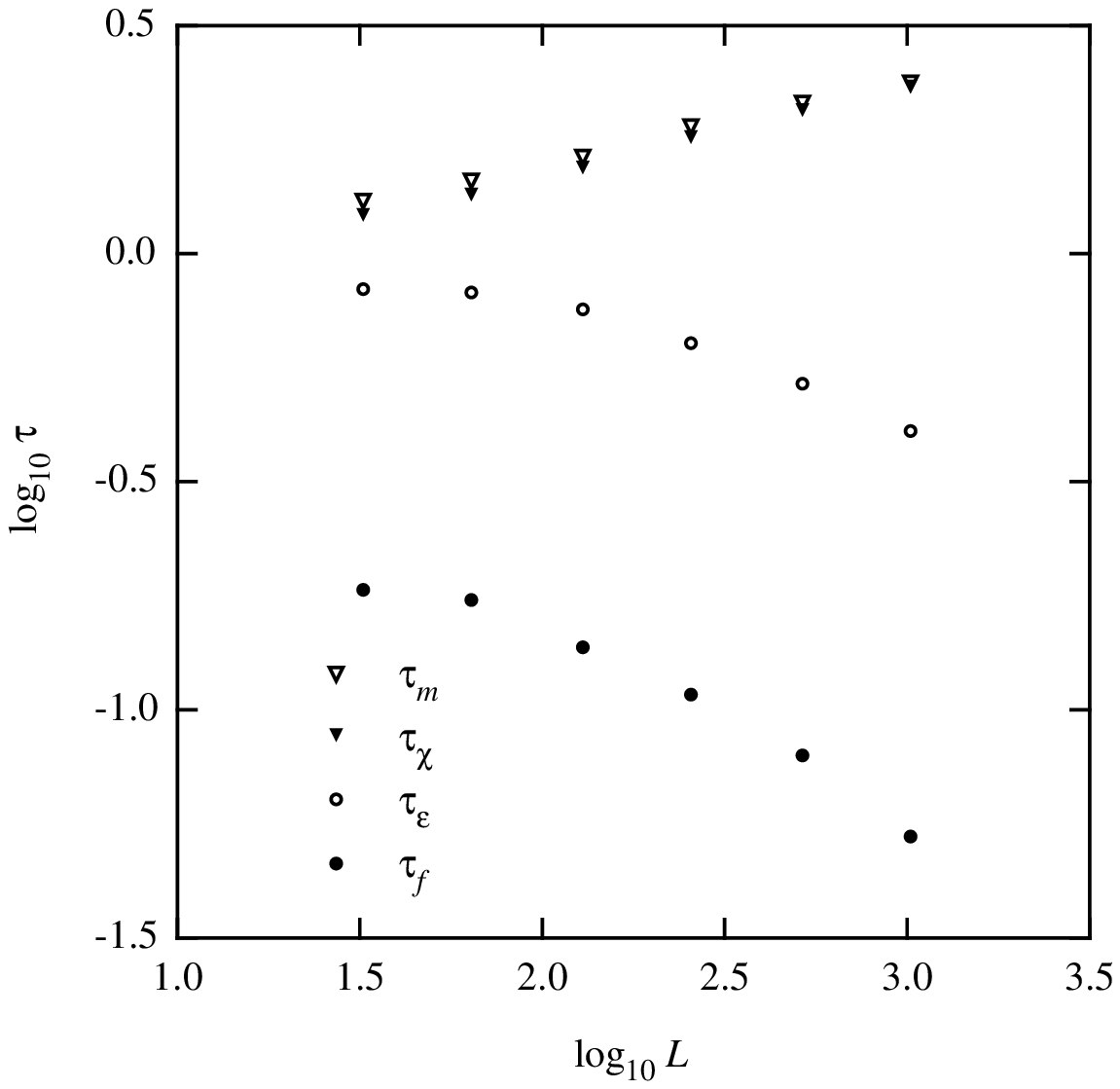}}
  \end{picture}
  \caption{$\log_{10} \tau$ vs.\ $\log_{10} L$ for the IC algorithm, where 
  $\tau$ is the integrated autocorrelation time and $32\le L \le 1024$ 
  is the lattice size, plotted for the $2d$ 3-state Potts model for the 
  energy per spin $\varepsilon$, the ratio $f$ of occupied to 
  satisfied bonds, the fraction $m$ of spins in the largest cluster, 
  and the susceptibility $\chi$.} 
  \figlabel{q3taus}
 \end{center}
\end{figure}

\newpage

\begin{figure}
 \begin{center}
  \begin{picture} (274, 462) (0,-63) 
   \put(0,0){\epsfbox[140 289 414 563]{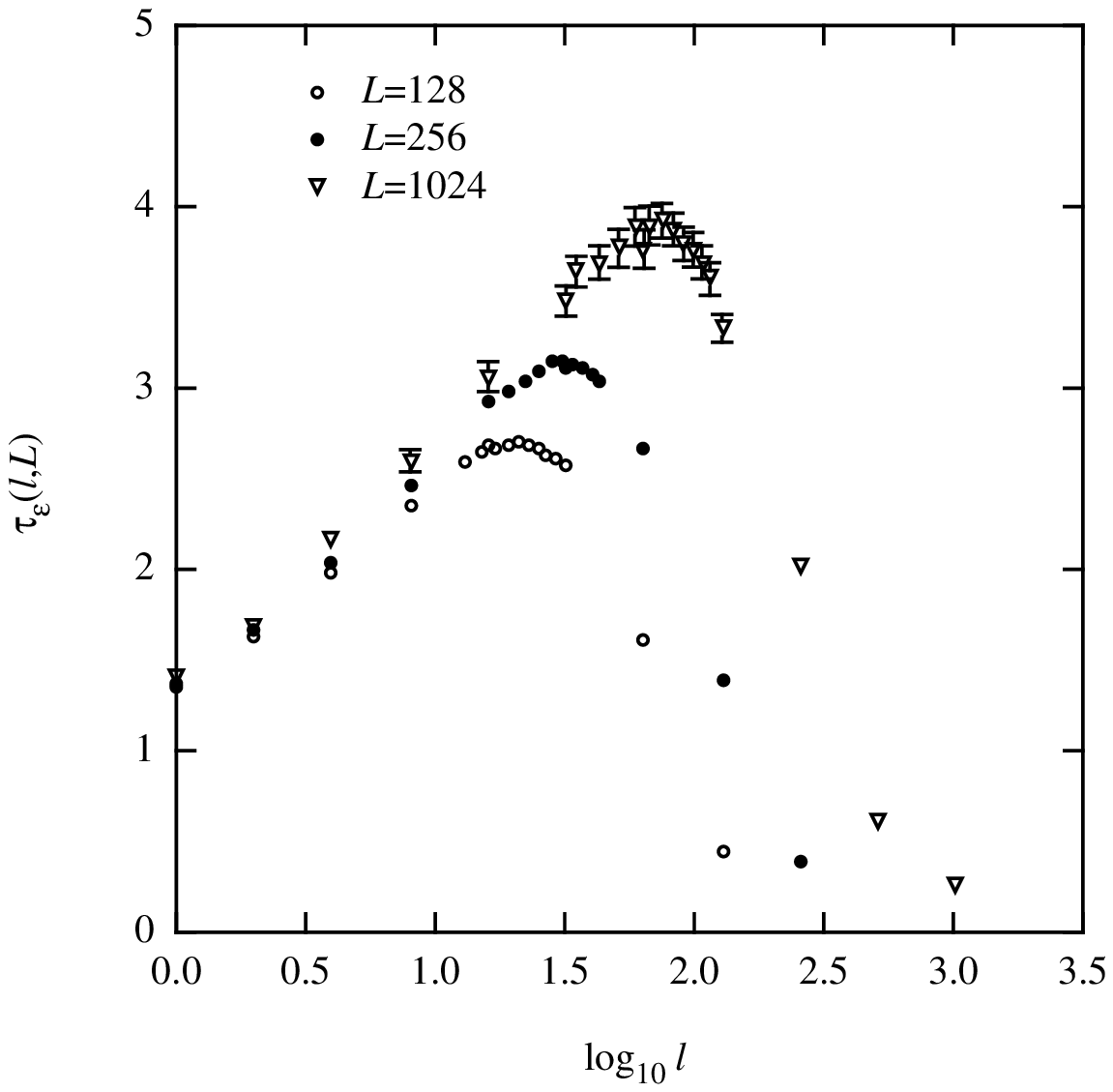}}
  \end{picture}
  \caption{The integrated autocorrelation time 
  $\tau_{\varepsilon}(l,L)$ for the energy per spin $\varepsilon(l,L)$ 
  measured on an $l\times l$ sub-lattice of an $L\times L$ lattice, 
  plotted vs.\ $\log_{10} l$ for $L=128, 256, 1024$ for the IC algorithm in the 
  case of the $2d$ Ising model.} 
  \figlabel{q2tauvsl}
 \end{center}
\end{figure}

\newpage

\begin{figure}
 \begin{center}
  \begin{picture} (274, 462) (0,-63)
   \put(0,0){\epsfbox[140 289 414 563]{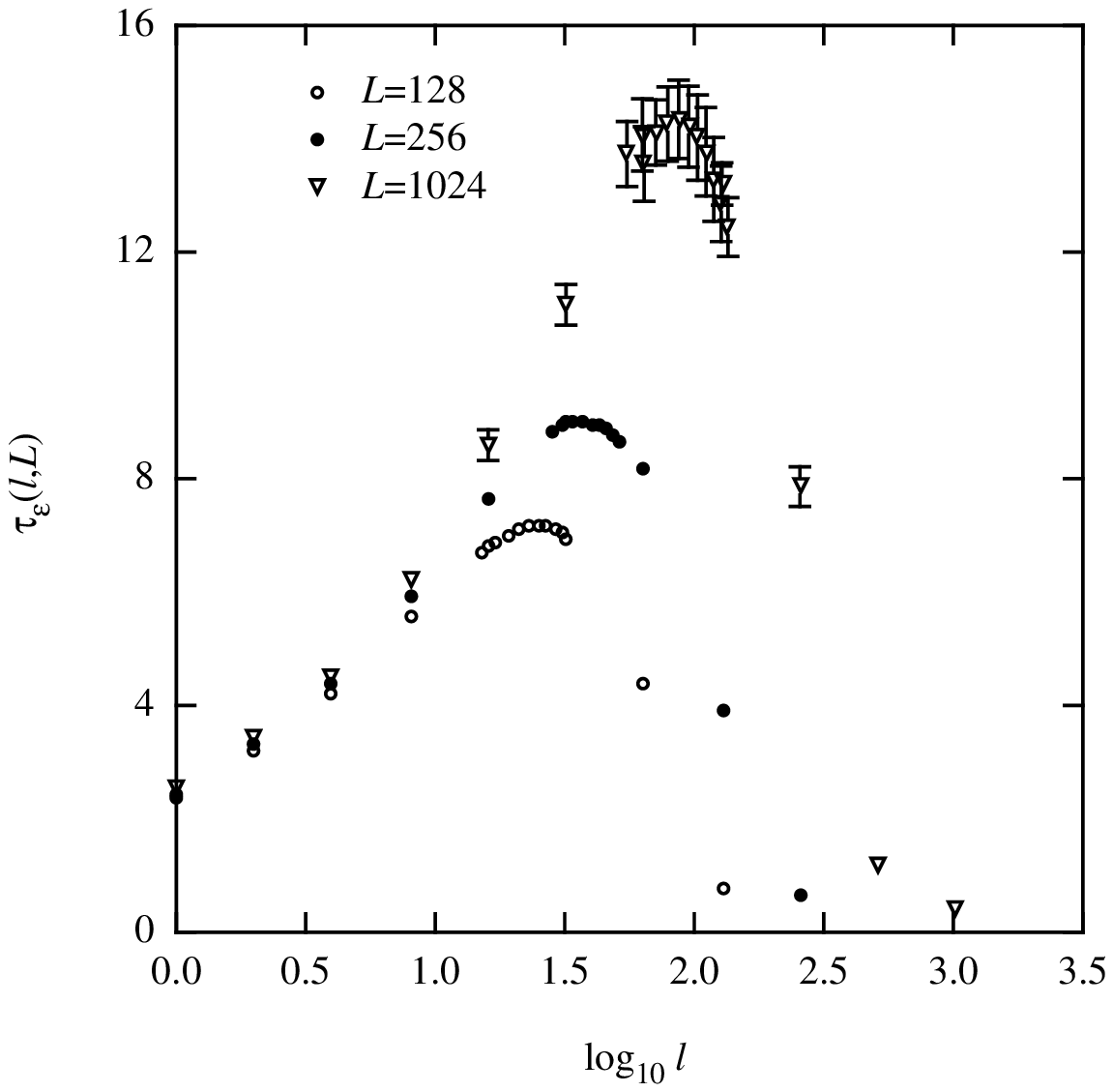}}
  \end{picture}
  \caption{The integrated autocorrelation time 
  $\tau_{\varepsilon}(l,L)$ for the energy per spin $\varepsilon(l,L)$ 
  measured on an $l\times l$ sub-lattice of an $L\times L$ lattice, 
  plotted vs.\ $\log_{10} l$ for $L=128, 256, 1024$ for the IC algorithm in the 
  case of the $2d$ 3-state Potts model.}
  \figlabel{q3tauvsl}
 \end{center}
\end{figure}

\newpage

\begin{figure}
 \begin{center}
  \begin{picture} (274, 462) (0,-63) 
   \put(0,0){\epsfbox[140 289 414 563]{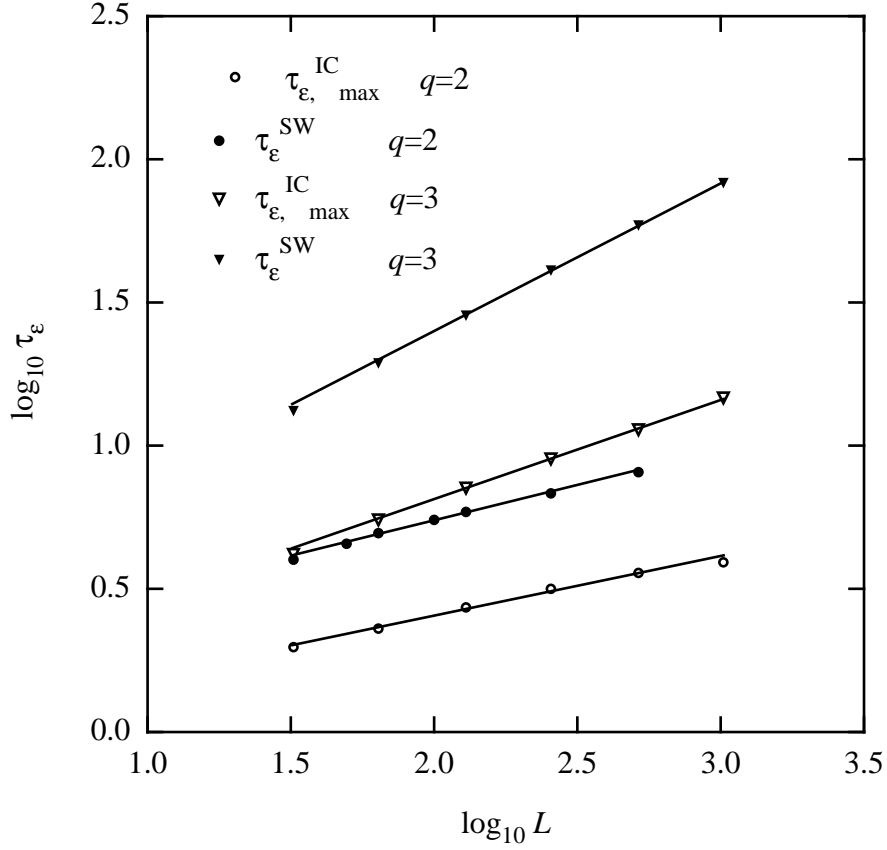}}
  \end{picture}
 \caption{$\log_{10} \tau_{\varepsilon}^{{\rm SW}}$ and $\log_{10} 
  \tau_{\varepsilon,{\rm max}}^{{\rm IC}}$ for the $2d$ Ising and 
  3-state Potts models.  $\tau_{\varepsilon}^{{\rm SW}}$ is the 
  integrated energy autocorrelation time for the Swendsen-Wang 
  algorithm on an $L\times L$ lattice and 
  $\tau_{\varepsilon,{\rm max}}^{{\rm IC}}$ is the height of the maximum 
  in Figs.~\ref{fig:q2tauvsl} and~\ref{fig:q3tauvsl}.  The solid lines 
  are linear fits to the data (see text for further details).}
  \figlabel{taumaxvsL}
 \end{center}
\end{figure}

\newpage

\begin{figure}
 \begin{center}
  \begin{picture} (274, 462) (0,-63) 
   \put(0,0){\epsfbox[140 289 414 563]{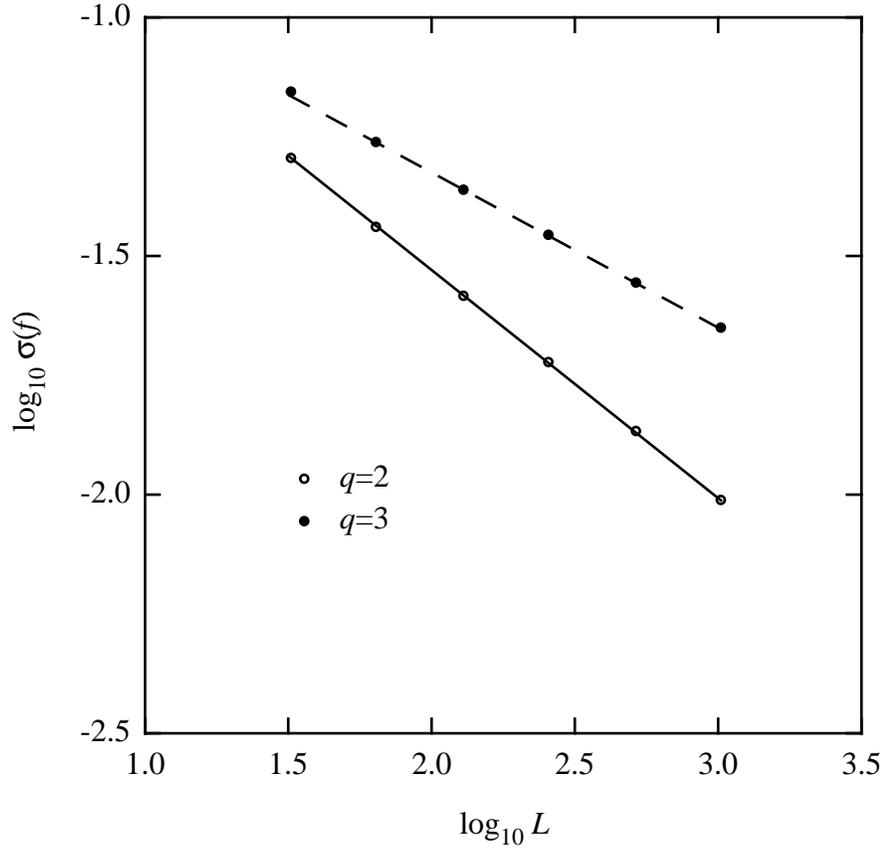}}
  \end{picture}
  \caption{$\log_{10} \sigma(f)$ vs.\ $\log_{10} L$ for the IC algorithm, plotted 
  for the $2d$ Ising and 3-state Potts models.  $\sigma(f)$ is the 
  standard deviation in the ratio $f$ of occupied to satisfied bonds
  and $L$ is the lattice size.  The solid (dashed) line is a 
  linear fit to the $q=2$ ($q=3$) data for $32\le L\le 1024$ ($128\le 
  L\le 1024$) and has slope $-.4781$ ($-.3252$).}
  \figlabel{sigfvsL}
 \end{center}
\end{figure}

\newpage

\begin{figure}
 \begin{center}
  \begin{picture} (274, 462) (0,-63) 
   \put(0,0){\epsfbox[140 289 414 563]{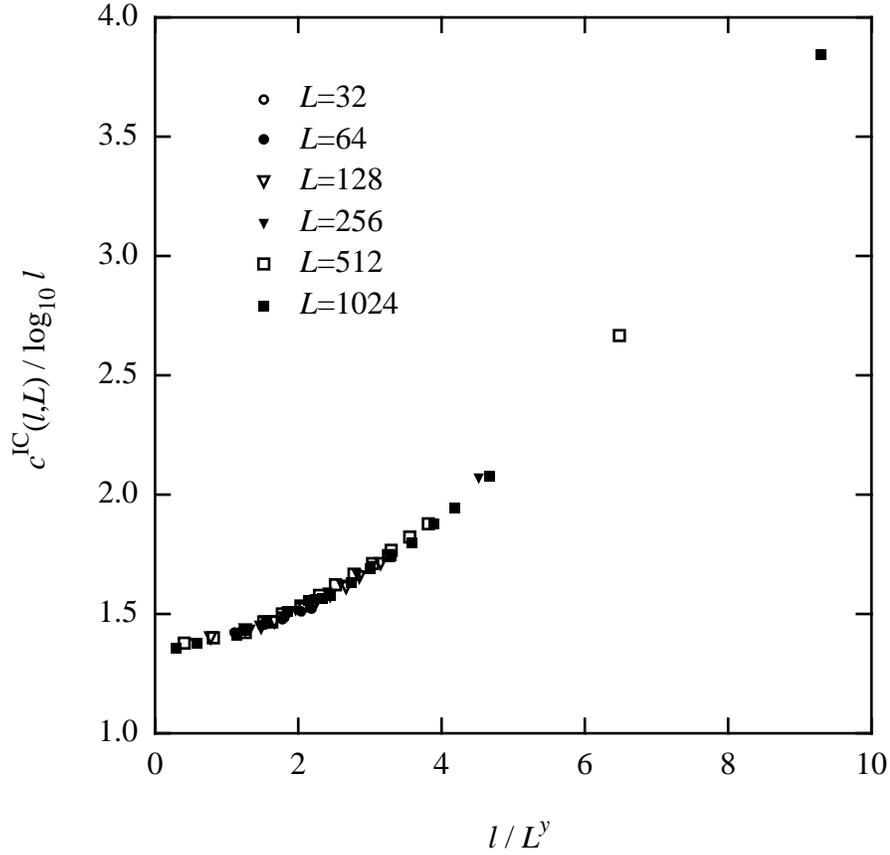}}
  \end{picture}
  \caption{$c^{{\rm IC}}(l,L)/\log_{10} l$ vs.\ $l/L^{y}$ for the IC 
  algorithm, plotted for the $2d$ Ising model for $32\le L \le 1024$ 
  and $8\le l\le L/4$.  Here, $c^{{\rm IC}}(l,L)\equiv 
  l^{d}{\rm var}(\varepsilon(l,L))$, where $\varepsilon(l,L)$ is the 
  energy per spin measured on an $l\times l$ sub-lattice of an 
  $L\times L$ lattice and $y=b\nu$, where $b=.4781$ is minus the slope of 
  the solid line in \fig{sigfvsL} and $\nu=1$ is the 
  correlation-length exponent.} 
  \figlabel{q2ccollapse}
 \end{center}
\end{figure}

\newpage

\begin{figure}
 \begin{center}
  \begin{picture} (274, 462) (0,-63) 
   \put(0,0){\epsfbox[140 289 414 563]{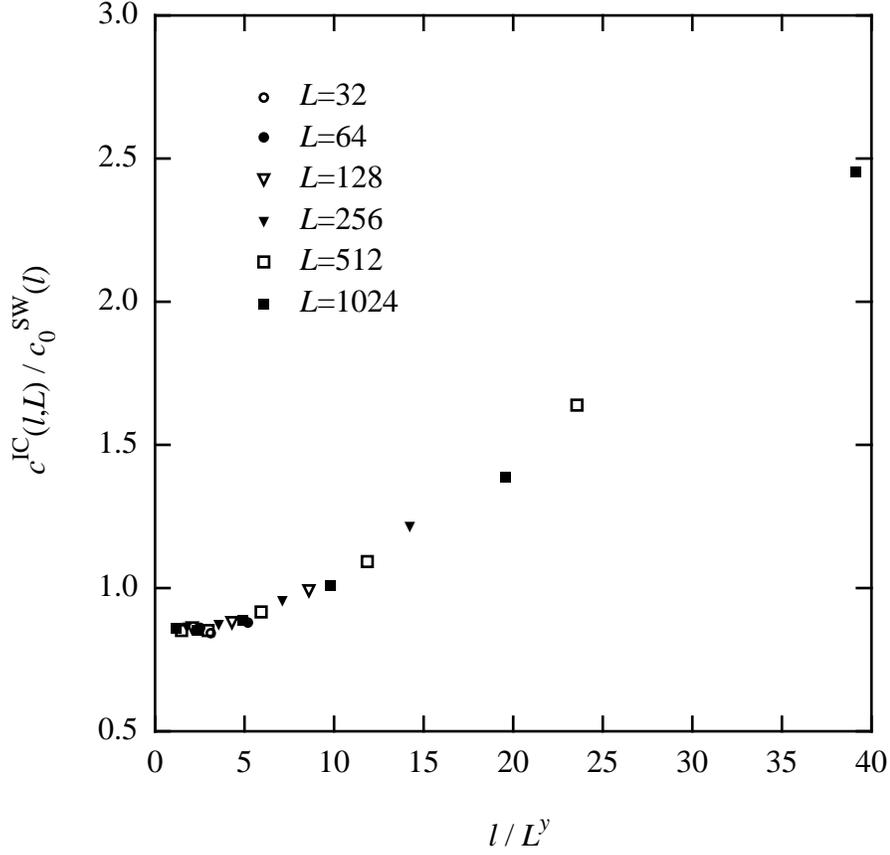}}
  \end{picture}
  \caption{$c^{{\rm IC}}(l,L)/c_{0}^{{\rm SW}}(l)$ vs.\ $l/L^{y}$ for the IC 
  algorithm, plotted for the $2d$ 3-state Potts model for $32\le L \le 1024$ 
  and $8\le l\le L/4$.  Here, $c^{{\rm IC}}(l,L)\equiv 
  l^{d}{\rm var}(\varepsilon(l,L))$, where $\varepsilon(l,L)$ is the 
  energy per spin measured on an $l\times l$ sub-lattice of an 
  $L\times L$ lattice, $c_{0}^{{\rm SW}}(l)$ is the specific heat for an 
  $l\times l$ lattice,
  $y=b\nu$ is the crossover exponent, $b=.3252$ is minus the slope of the dashed line in 
  \fig{sigfvsL} and $\nu=5/6$ is the correlation-length exponent.} 
  \figlabel{q3ccollapse}
 \end{center}
\end{figure}

\newpage

\begin{figure}
\begin{center}
  \begin{picture} (274, 462) (0,-63)
   \put(0,0){\epsfbox[140 289 414 563]{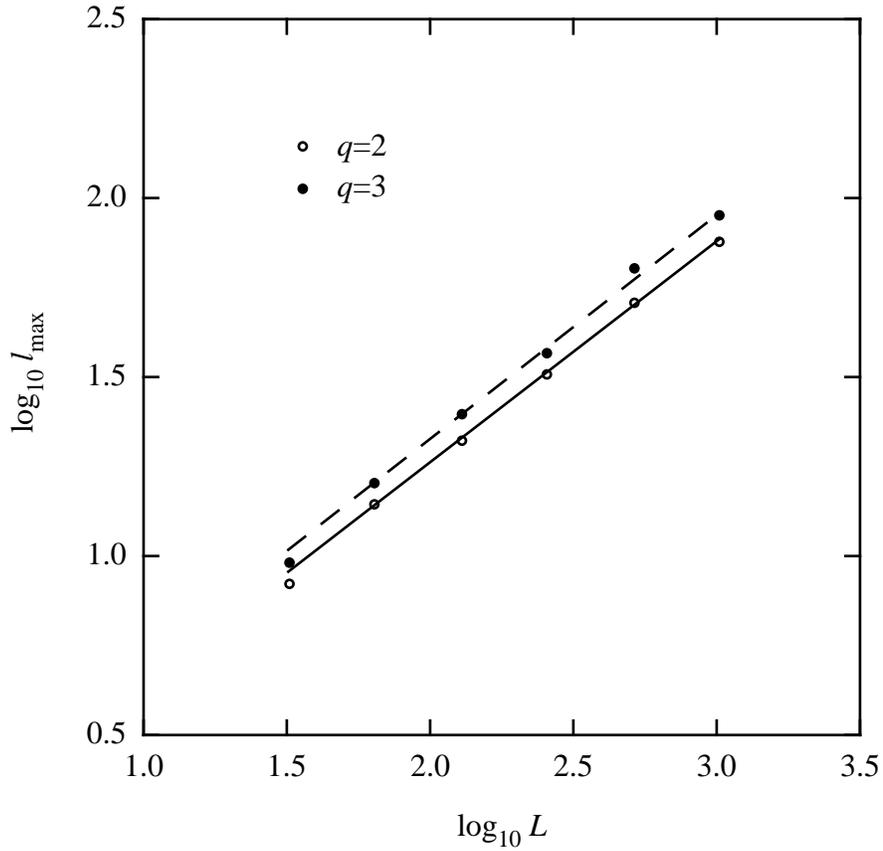}}
  \end{picture}
  \caption{$\log_{10} l_{{\rm max}}$ vs.\ $\log_{10} L$ for the IC algorithm, 
  plotted for the $2d$ Ising and 3-state Potts models.  
  $l_{{\rm max}}$ is the location of the maximum
  in Figs.~\ref{fig:q2tauvsl} and~\ref{fig:q3tauvsl}, 
  and $L$ is the 
  lattice size.  The solid (dashed) line is a linear fit to the 
  $q=2$ ($q=3$) data for $64\le L \le 1024$ and has slope .6176 (.626).} 
  \figlabel{lmaxvsL}
  \end{center}
\end{figure}

\newpage

\begin{figure}
 \begin{center}
  \begin{picture} (274, 462) (0,-63) 
   \put(0,0){\epsfbox[140 289 414 563]{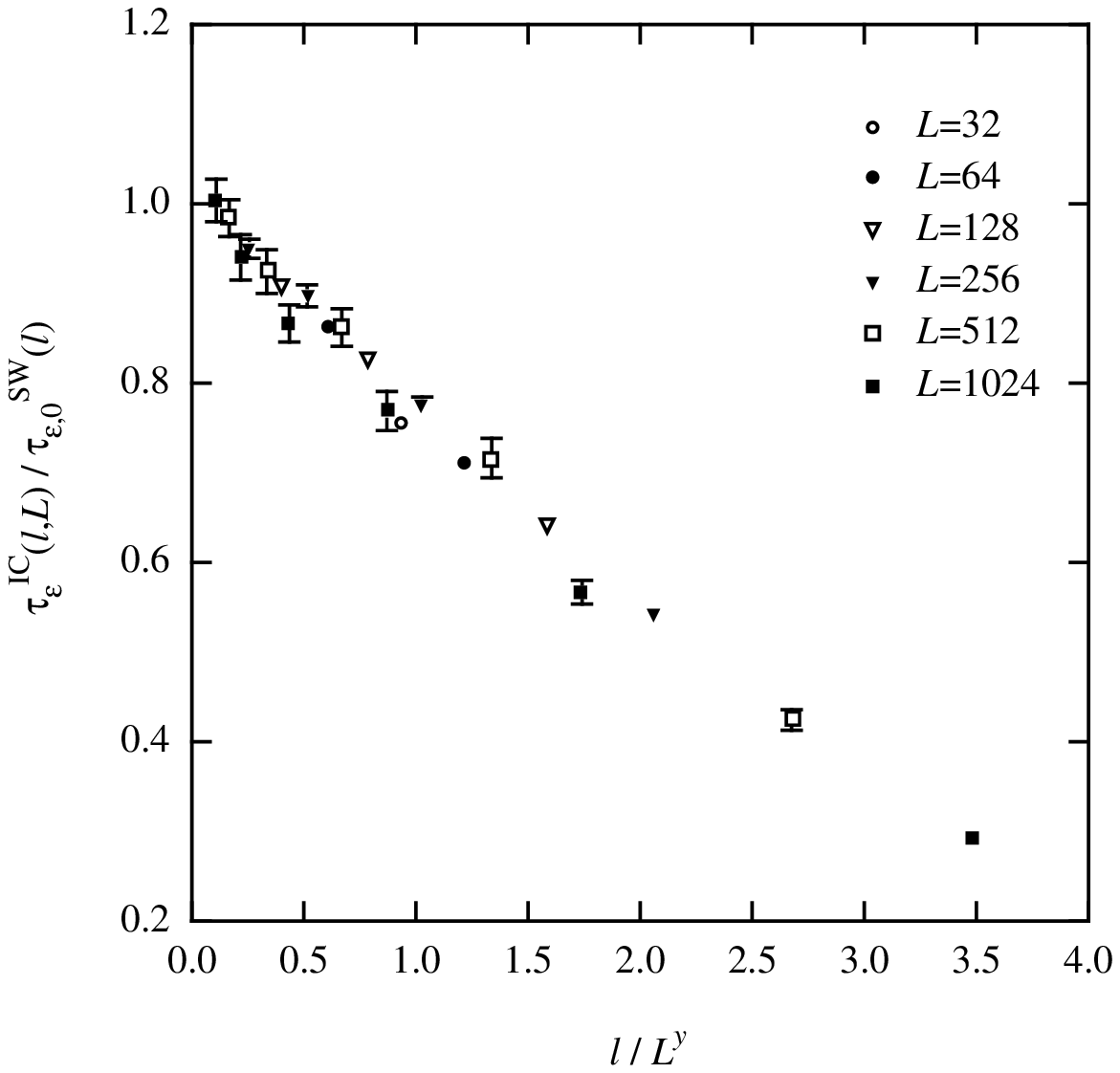}}
  \end{picture}  
  \caption{$\tau_{\varepsilon}^{{\rm IC}}(l,L)/\tau_{\varepsilon,0}^{{\rm SW}}(l)$ 
  vs.\ $l/L^{y}$ for the $2d$ Ising model for $32 \le L\le 1024$ 
  and $8\le l\le L/4$.  $\tau_{\varepsilon}^{{\rm IC}}(l,L)$ is the 
  quantity plotted in \fig{q2tauvsl}, 
  $\tau_{\varepsilon,0}^{{\rm SW}}(l)$ is the integrated energy 
  autocorrelation time for the Swendsen-Wang algorithm on an $l\times 
  l$ lattice and $y=.6176$ is the slope of the solid line in \fig{lmaxvsL}.} 
  \figlabel{q2taucollapse}
 \end{center}
\end{figure}

\newpage

\begin{figure}
 \begin{center}
  \begin{picture} (274, 462) (0,-63) 
   \put(0,0){\epsfbox[140 289 414 563]{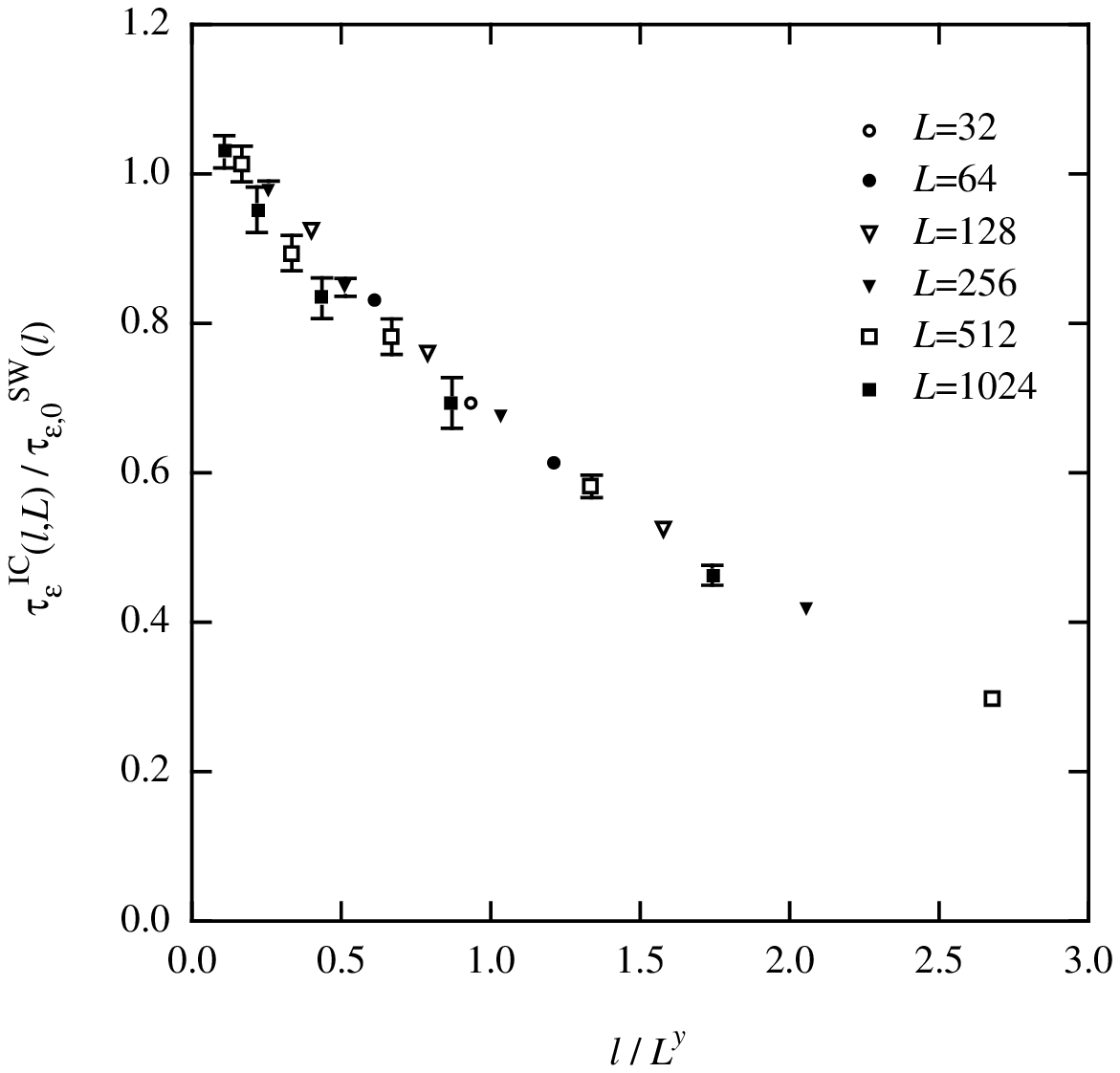}}
  \end{picture}
  \caption{$\tau_{\varepsilon}^{{\rm IC}}(l,L)/\tau_{\varepsilon,0}^{{\rm SW}}(l)$ 
  vs.\ $l/L^{y}$ for the $2d$ 3-state Potts model for $32 \le L\le 1024$ 
  and $8\le l\le L/4$.  $\tau_{\varepsilon}^{{\rm IC}}(l,L)$ is the 
  quantity plotted in \fig{q3tauvsl}, 
  $\tau_{\varepsilon,0}^{{\rm SW}}(l)$ is the integrated energy 
  autocorrelation time for the Swendsen-Wang algorithm on an $l\times 
  l$ lattice and $y=.626$ is the slope of the dashed line in \fig{lmaxvsL}.} 
  \figlabel{q3taucollapse}
 \end{center}
\end{figure}

\end{document}